\documentclass[aps,prl,twocolumn,superscriptaddress,longbibliography,footnote]{revtex4-1}
\usepackage{bm}
\usepackage{mathrsfs,amssymb,amsmath,mathtools,amsthm,amsfonts}
\usepackage{graphicx}
\usepackage[hidelinks,colorlinks=true,linkcolor=blue,citecolor=blue,anchorcolor=blue,filecolor=blue,menucolor=blue,runcolor=blue,urlcolor=blue]{hyperref}
\usepackage{color}

\begin{document}
\title{Index Theorem and Vortex Kinetics in Bose-Einstein Condensates on a Haldane Sphere with a Magnetic Monopole}
\author{Xi-Yu Chen}
\email{These authors contribute equally to this paper.}
\affiliation{Institute of Modern Physics and school of Physics, Northwest University, Xi'an 710127, China}
\affiliation{Shaanxi Key Laboratory for Theoretical Physics Frontiers, Xi'an 710127, China}
\affiliation{Peng Huanwu Center for Fundamental Theory, Xi'an 710127, China}
\author{Lijia Jiang}
\email{These authors contribute equally to this paper.}
\affiliation{Institute of Modern Physics and school of Physics, Northwest University, Xi'an 710127, China}
\affiliation{Shaanxi Key Laboratory for Theoretical Physics Frontiers, Xi'an 710127, China}
\affiliation{Peng Huanwu Center for Fundamental Theory, Xi'an 710127, China}
\author{Tao Yang}
\email{yangt@nwu.edu.cn}
\affiliation{Institute of Modern Physics and school of Physics, Northwest University, Xi'an 710127, China}
\affiliation{Shaanxi Key Laboratory for Theoretical Physics Frontiers, Xi'an 710127, China}
\affiliation{Peng Huanwu Center for Fundamental Theory, Xi'an 710127, China}
\author{Jun-Hui Zheng}
\email{junhui.zheng@nwu.edu.cn}
\affiliation{Institute of Modern Physics and school of Physics, Northwest University, Xi'an 710127, China}
\affiliation{Shaanxi Key Laboratory for Theoretical Physics Frontiers, Xi'an 710127, China}
\affiliation{Peng Huanwu Center for Fundamental Theory, Xi'an 710127, China}

\date{\today}

\begin{abstract}
The geometry-gauge interplay constitutes a fundamental issue in quantum physics, with profound implications spanning from quantum gravity to topological matter. Here, we investigate the dynamic effects of geometry-gauge interplay in Bose-Einstein condensates (BECs) on a Haldane sphere with a magnetic monopole. We reveal an index theorem that establishes a correspondence between BEC vortices and the topology of the gauge field, enabling the construction of vortex-monopole composites. Furthermore, we derive the universal logarithmic interaction between composites, which governs the structure of the ground-state vortex lattice. By developing a kinetic theory, we predict scale-invariant vortex dynamics and an emergent duality. Both are confirmed through numerical simulations. This work first presents the dynamical coupling mechanism between spatial geometry and gauge fields, providing deep insights into superfluid systems with topological gauge structures in curved space.
\end{abstract}

\maketitle

Quantum phenomena intrinsically arise from coherent superpositions of phase factors contributed by all equal-weight paths \cite{Feynman1948}. The relative phases among different paths depend on both the underlying geometric structure (such as curvature and topology) and gauge fields (for instance, the Aharonov-Bohm effect) \cite{Batelaan2009}. The geometry-gauge interplay has been a significant issue. It is involved in quantum gravity that unifies gravity (geometrized in general relativity) and gauge interactions \cite{Carlip2001}, quantum field theory in curved space or non-inertial systems \cite{Parker2009}, and topological phenomena in condensed matter systems under curved real or momentum space \cite{Huertas-Hernando2006,Haldane2004}.

The Haldane sphere system, featuring constant spatial curvature and monopole-induced uniform magnetic field, is an ideal platform for investigating emergent topological phases in many-body systems. It is particularly beneficial for exploring the synergy between geometry and gauge fields due to its few parameters and intrinsic high [SU(2)] symmetry. Previous studies have demonstrated that the non-relativistic scalar fermions and Dirac spinor fields on the Haldane sphere exhibit spherical quantum Hall effects or fractional quantization \cite{Haldane1983,Fano1986,Greiter2018,Dolan2020,Hsiao2020}.
Yet, systematic investigations of scalar, spinor, and vector bosons on the Haldane sphere remain lacking.

A salient feature of bosonic systems is that they are not subject to the Pauli exclusion principle, and hence can support Bose-Einstein condensates (BECs). Heretofore, both the simulation of magnetic monopoles \cite{Ray2014,Ray2015,Pietil2009} and the creation of bubble-shaped systems \cite{Carollo2022,Jia2022} have been achieved in ultracold atomic BEC experiments. A scheme of simulating half-integer-type monopoles has also been proposed \cite{Zhou2018,Zheng2025}. These advancements pave the way to realize diverse Haldane spheres in BECs. Moreover, prior studies of monopole-free spherical BECs, including quantum phase transitions, supersolidity, Berezinskii-Kosterlitz-Thouless (BKT) physics, and quantum turbulence \cite{Tononi2020,Ciardi2024,Tononi2019,Tononi2022,Kanai2021,Yu2025,Dritschel2015}, have established crucial benchmarks for exploring gauge field effects in such a curved space. So far, although the ground-state properties of the BEC on the Haldane sphere have been well investigated \cite{Zhou2018,Ruokokoski2011}, how quantum coherence under geometry-gauge interplay influences nonequilibrium dynamics of the BEC remains essentially unexplored.

In this Letter, we explore the BEC dynamics on the Haldane sphere. We establish a vortex-monopole index theorem that relates the total vortex winding number to the magnetic charge of the monopole in the sphere. This theorem enables us to decompose the single monopole and construct composite topological particles via binding each vortex with a matching monopole. Each composite exhibits a quantized angular momentum (aligned with its vortex core position vector), generalizing composite particle theories in fractional quantum Hall systems \cite{Tsui1982,Laughlin1983,Jain1989, Halperin1993,Kang1993,Kamilla1996,Sitko1997,Pan2003}.

When interatomic interactions dominate, we show that these composites obtain a current-mediated interaction which determines the ground-state vortex lattice configuration. We formulate a kinetic theory for vortices, revealing that the precession of composites --- driven by their mutual interactions --- fully dictates the vortex trajectories. This theory extends kinetic theory of point vortices on monopole-free curved space \cite{Turner2010,Yu2024,Sun2020,Bereta2021,Kanai2021,Bogomolov1977} to systems with monopole fields. Intriguingly, the vortex motion depends solely on the initial vortex distribution, independent of any other system parameters. Furthermore, we uncover a duality between vortex-vortex and vortex-antivortex systems, linking Haldane systems with different monopole charges.

The analytical predictions are confirmed by numerical simulations using the spherical Gross-Pitaevskii equation (GPE). We elucidate a dynamical interplay mechanism between spatial geometry and gauge fields, offering deep insights into superfluid systems on curved spaces with topological gauge structures.

{\it Gauge Transformation.} Far below the BEC transition temperature, the BEC dynamics on a Haldane sphere with radius $R$ is described by the GPE ($\hbar =1$),
\begin{equation}\label{gpe}
i  \partial_t \Psi_{\textbf n} = \frac{ (-i  \nabla_s - \textbf{A}_{\textbf n})^{2}}{2m} \Psi_{\textbf n} + \lambda ~|\Psi_{{\textbf n}}|^2 \Psi_{{\textbf n}},
\end{equation}
where $\nabla_s = (1/R) [\hat{\textbf e}_\theta \partial_\theta + (1/\sin\theta) \hat{\textbf e}_\phi \partial_\phi]$ is the surface gradient operator, $\textbf{A}_{\textbf n}$ is the gauge field of the magnetic monopole, and $\lambda$ is the interatomic interaction strength. The magnetic field has a source, hence a singularity line (Dirac string) exists in $\textbf{A}_{\textbf n}$. For Dirac string  along any direction $\textbf n = (\sin\theta_{\textbf n} \cos\phi_{\textbf n}, \sin\theta_{\textbf n} \sin\phi_{\textbf n}, \cos\theta_{\textbf n})$, we have
\begin{equation}\label{gauge}
\textbf{A}_{\textbf n}(\textbf r) = \frac{-g  \textbf n \times \textbf r}{r (r - \textbf n \cdot \textbf r)},
\end{equation}
where $g$ is the magnetic charge and $\textbf n$ identifies a gauge choice. The magnetic field $\textbf{B} = \nabla \times \textbf{A}_{\textbf n} = g  \textbf r/ r^3$ is gauge invariant. In regions excluding singular points, the wavefunctions transform as
$\Psi_{\textbf m} (\textbf r) = e^{i \xi_{\textbf m \textbf n}(\textbf r)}\Psi_{ \textbf n}(\textbf r)$ between two gauges $\textbf m$ and $\textbf n$. The real function $\xi_{\textbf m \textbf n}(\textbf r)$ satisfies $\nabla_s \xi_{\textbf m \textbf n}(\textbf r) = \textbf{A}_{\textbf m}(\textbf r) - \textbf{A}_{\textbf n}(\textbf r)$ and obeys the cocycle conditions $\xi_{\textbf m \textbf n}(\textbf r) = \xi_{\textbf m \textbf l}(\textbf r) + \xi_{\textbf l \textbf n}(\textbf r)$ and $\xi_{\textbf m \textbf n}(\textbf r) =- \xi_{\textbf n \textbf m}(\textbf r)$ for gauges $\textbf l$, $\textbf m$, and $\textbf n$. As demonstrated in the Supplementary Material (SM) \cite{jhzheng2025}, the explicit form of the transformation $\xi_{\textbf n \bar{\textbf z}}(\textbf r)$ can be obtained by selecting an integral path, where $\bar{\textbf z} = -\textbf z \equiv (0,0,-1)$ corresponds to the south-pole gauge. This leads to the general transformation $\xi_{\textbf m\textbf n} = \xi_{\textbf m \bar{\textbf z}} - \xi_{\textbf n\bar{\textbf z}}$, which naturally extends the Wu-Yang (north-south) transformation \cite{Yang1975}.

{\it Vortex-Monopole Index Theorem.} GPE \eqref{gpe} possesses rotational symmetry with generators being the angular momentum operators: $ {\hat{ \textbf L}}(g) = \textbf{r} \times {\hat{ \textbf p}} - g {\textbf{r}}/{r}$. In the $\bar{\textbf z}$-gauge, the kinetic momentum is ${\hat{\textbf p}}(g) = -i\nabla - \textbf{A}_{\bar{\textbf z}}$ \cite{Yang1976}. The formation of vortices spontaneously breaks the symmetry. Let integer $c'$ denote the total winding number of vortices on the sphere {\it excluding} the south pole: $c' = \sum'_i (1/2\pi)\oint \nabla \arg[\Psi_{\bar {\textbf z}}] \cdot d \bm \ell$, where the loop integral (right-handed) surrounds density zeros and the prime excludes the south pole. Then, a right-handed loop integral surrounds the south pole yields a phase accumulation of $-2\pi c'$. After Wu-Yang transformation to the ${\textbf z}$-gauge, where both $ \textbf{A}_{{\textbf z}}$ and $ \Psi_{{\textbf z}}$ are well-defined at the south pole \cite{Yang1975}, the accumulation becomes $2\pi(2g - c')$, corresponding to a winding number $2g - c'$ at the south pole. Consequently, the {\it total} vortex winding number on the full sphere is exactly
\begin{equation}\label{winding}
    2 g = \sum_i c_i,
\end{equation}
where $\{c_1, c_2, \cdots\}$ is the vortex winding number distribution on sphere. This theorem builds a relation between topological defects of a scalar field on a sphere and the first Chern number of the background magnetic field, giving a constraint on vortex distributions. The theorem is an application of Atiyah-Singer index theorem \cite{Atiyah1963,Dolan2020} and consistent with Dirac quantization (\(g \in \mathbb{Z}/2\)).

{\it Vortex-Monopole Composite.} If only one vortex forms on the sphere, the winding number is $c = 2g$ according to the index theorem. The BEC wavefunction for such a single vortex-monopole composite structure reads
\begin{equation}\label{composite}
\Psi_{\textbf n} = \sqrt{\rho} \, \exp\left[i \Phi^c_{\textbf n}(\textbf{r},\textbf n_v)\right],
\end{equation}
where $\textbf n_v$ is the unit vector pointing to the vortex center. In the regime where interatomic interaction dominates, $ \mu = \lambda \rho \gg \varepsilon_0 = 1/m R^2$, the healing length $\xi = 1 / \sqrt{2  m \mu}$ is much smaller than $R$. Thus the vortex can be treated as a point defect, since the vortex core size is about $ |c| \xi \ll R$ \cite{Pethick2008}. The BEC density becomes nearly uniform: $\rho \simeq  N/4 \pi R^2$ except the defects. In the $\bar{\textbf{z}}$-gauge, for a vortex at the north pole, $\Phi^c_{\bar{\textbf z}}(\textbf{r}, \textbf{z}) = c \phi$ \cite{Zheng2025}. Similarly, for an arbitrary vortex position $R \textbf{n}_v$, the phase in the gauge $\bar{\textbf{n}}_v \equiv -{\textbf{n}}_v$ takes the form $\Phi^c_{\bar{\textbf n}_v}(\textbf{r}, \textbf{n}_v) = c \phi'$, where $\phi'$ is the azimuthal angle in a rotated coordinate frame with ${\textbf{n}}_v$ as the $z'$-axis. Transforming back to the $\bar{\textbf{z}}$-gauge yields
\begin{equation}
  \Phi^c_{\bar{\textbf z}}(\textbf{r}, \textbf{n}_v) = c \phi' - \xi_{\bar{\textbf n}_v \bar{\textbf z}}(\textbf{r}).
\end{equation}
The velocity field
\begin{equation}
\textbf{v}_c(\textbf{r}; \textbf{n}_v) \equiv \frac{\nabla \Phi^c_{\bar{\textbf n}_v}- \textbf{A}_{\bar{\textbf n}_v}}{m} = \frac{c}{2 m R} \frac{\textbf{n}_v \times \textbf{r}}{R - \textbf{n}_v \cdot \textbf{r}},
\end{equation}
diverges at the vortex center and the vorticity is $\nabla \times \textbf{v}_c = -{\textbf{B}}/{m}$ elsewhere \cite{jhzheng2025}. The angular momentum per atom, $\bm{l} = (1/4\pi) \int d\Omega \left[\textbf r \times m \textbf{v}_c - g \textbf r/ r \right]= {c} \textbf{n}_v /2$, is quantized \cite{Zheng2025}. Thus, the composite can be equivalent to a spinful particle with spin (anti)parallel with the vortex position. Both $\textbf{v}_c$ and $\bm{l}$ are gauge-invariant and $\lambda$-independent.

{\it Composite-Composite Interaction.} For a system with an arbitrary vortex distribution, we treat the monopole --- based on the index theorem --- as an overlapping of monopoles with magnetic charges $g_i = c_i/2$, i.e., $\textbf{A}_{\textbf n}(g) = \sum_i \textbf{A}_{\textbf n}({g_i})$. When the interatomic interaction dominates, the wavefunction can be written as
\begin{equation}\label{twovort}
\Psi_{\textbf n}(\theta,\phi) = \sqrt{\rho}\exp\Big[i \sum_i \Phi^{c_i}_{\textbf n}(\textbf r, \textbf n_{v}^i) \Big],
\end{equation}
where $\Phi^{c_i}_{\textbf n}(\textbf r, \textbf n_{v}^i)$ is the phase function of a single vortex-monopole composite with vortex winding number $c_i$ and monopole charge $g_i$, and $\rho$ is approximately uniform. A typical velocity distribution is plotted in the end matter.

The total energy of the system is $ R^2 \int d\Omega \big\{ ({\rho}/{2m}) |\big[-i\nabla_s - \textbf{A}_{\textbf n}(g)\big]  \exp[i \sum_i \Phi^{c_i}_{\textbf n}] |^{2} + {\lambda}\rho^2/2 \big\}$, where the interactomic interaction contributes a constant term. The kinetic energy simplifies to $({{\rho} m R^{2}}/{2}) \int d\Omega ~\big[ \sum_{i} \textbf v_{c_i}(\textbf r, \textbf n_{v}^i) \cdot \textbf v_{c_i}(\textbf r, \textbf n_{v}^i) + \sum_{i \neq j} \textbf v_{c_i}(\textbf r, \textbf n_{v}^i) \cdot \textbf v_{c_j} (\textbf r, \textbf n_{v}^j) \big]$, where the first term represents the {\it internal energy} of each composite (independent of the vortex positions) and the second term is the current-mediated interaction between composites. Although the cross term is formally similar to the interaction between point vortices on an {\it open} curved surface without magnetic fields \cite{Turner2010}, the velocity fields have been modified by the presence of the magnetic field. We demonstrate that the interaction between composites takes the logarithmic form,
\begin{equation}
E_{\text{int}} = -  \frac{N \varepsilon_0}{4} \sum_{i<j} c_i c_j \ln(1 - \textbf{n}_v^i \cdot \textbf{n}_v^j),
\label{eq:interaction}
\end{equation}
where $N$ is the total atom number \cite{jhzheng2025} (The interaction \eqref{eq:interaction} is different from that in the weak-interatom-interaction case, where the vortex-vortex interaction energy arises mainly from the interatomic interaction correction and is proportional to $N^2$ \cite{Zhou2018}). The interaction among composites does not explicitly rely on the magnetic field. Indeed, the physical effect of changing the monopole charge is via adjusting the vortex number distribution according to the index theorem. Moreover, $E_{\text{int}}$ is formally identical to the conventional form of vortex-vortex interaction on a regular sphere without a monopole. In the latter case, the system energy is evaluated using the Green's function method for {\it closed} manifolds, but there is no unique way to split the energy into single
particle energies and two-particle interaction energies since $\sum_i c_i =0$ \cite{Kanai2021,Bogomolov1977,Turner2010}. In our frame, the regular sphere corresponds to the special case with $g = 0$, where component monopoles with magnetic charge $g_i$ can be treated as auxiliary to construct the composite, providing a special way to split the energy. As a result, any spherical vortex system can be uniformly depicted by interacting composites, each of which comprises two classical topological solitons: a vortex and a matched monopole (significantly differs from the composite particle in the fractional quantum Hall effect \cite{Ezawa2008, Tsui1982,Laughlin1983,Jain1989, Halperin1993,Kang1993,Kamilla1996,Sitko1997,Pan2003}).

From Eq.\,\eqref{eq:interaction}, we conclude that vortices with the same sign prefer to be away from each other, as this lowers the interaction energy, while vortices with opposite signs are attractive. This further indicates that vortices with higher winding numbers are not energetically favorable, which may decay and go beyond a rigid vortex-monopole composite picture.

In the ground state, neither antivortices nor vortices with high winding numbers exist, and the total $2g$ ($g>0$) vortices of $c_i =1$ remain static. This is consistent with the relation between vortex density and magnetic flux density in the planar Abrikosov vortex lattice \cite{Cooper2001,Cooper2005,Schweikhard2004,Dingping2010}. $E_{\text{int}}$ uniquely determines the vortex lattice configuration, which is independent of system parameters such as mass, atom number, interatomic interaction strength, and sphere radius. The vortex lattice structures derived from Eq.\,\eqref{eq:interaction} are in accord with the GPE simulations, which are presented in the end matter.

{\it Vortex Kinetic Theory.} The dynamic wavefunction in the $\bar{\textbf z}$-gauge has the following form
\begin{equation}
  \Psi_{\bar{\textbf {z}}}(\textbf{r},t) =  \exp(-i \mathcal{E} t) \sqrt{\rho}  \prod_{j} \big[ \hat{\mathcal{R}}(g_j;\theta_v^j,\phi_v^j) \exp(i c_j \phi )\big],
\end{equation}
where $\hat{\mathcal{R}}(g_j;\theta_v^j,\phi_v^j) = \exp[-i\hat{L}_{z}(g_j) \phi_v^j] \exp[-i\hat{L}_{y}(g_j) \theta_v^j]$ rotates a vortex from the north pole to the real-time position $(\theta_v^j(t),\phi_v^j(t))$ and $\exp(-i \mathcal{E} t)$ is a dynamic phase. The Lagrangian $\mathcal{L} =  R^2 \int d\Omega (i\Psi^*_{\bar{\textbf {z}}} \partial_t \Psi_{\bar{\textbf {z}}}) - E_{\text{int}}$ becomes a function of vortex core positions and their velocities \cite{jhzheng2025}:
\begin{equation}
\mathcal{L} (\theta_v^1,\phi_v^1, \cdots, \dot{\theta}_v^1, \dot{\phi}_v^1, \cdots) = N \sum_i \frac{c_i}{2} \dot{\phi}_v^i\cos\theta_v^i - E_{\text{int}},
\end{equation}
The Euler-Lagrange equation gives the precession of the spins ($\bm l_ i \equiv c_i \textbf{n}^i_v /2$) of composites \cite{jhzheng2025}
\begin{equation}
\dot{\bm{l}}_i = \bm{\omega}_i \times \bm{l}_i, \quad \bm{\omega}_i = \frac{\nabla_{\bm{l}_i} E_{\text{int}}}{N}.
\label{eom}
\end{equation}
This means that, with the assumption of a rigid unit of vortex-monopole composite structure, the dynamics of each composite (or the vortex trajectory) is depicted by the precession of the spin under a torque contributed by the composite-composite interaction. This equation of motion unifies the theory of point-vortex dynamics on a sphere with arbitrary magnetic charge ~\cite{Turner2010,Yu2024,Sun2020,Bereta2021}. Moreover, since $E_{\text{int}} \propto N \varepsilon_0$, both dimensionless precession rates \( |\bm{\omega}_i t_0| \) (where \( t_0 \equiv \varepsilon_0^{-1} \)) and spatial trajectories --- depend only on the initial vortex configuration specified by positions and winding numbers.

\begin{figure*}[t]
\centering
\includegraphics[angle=0,width=\textwidth]{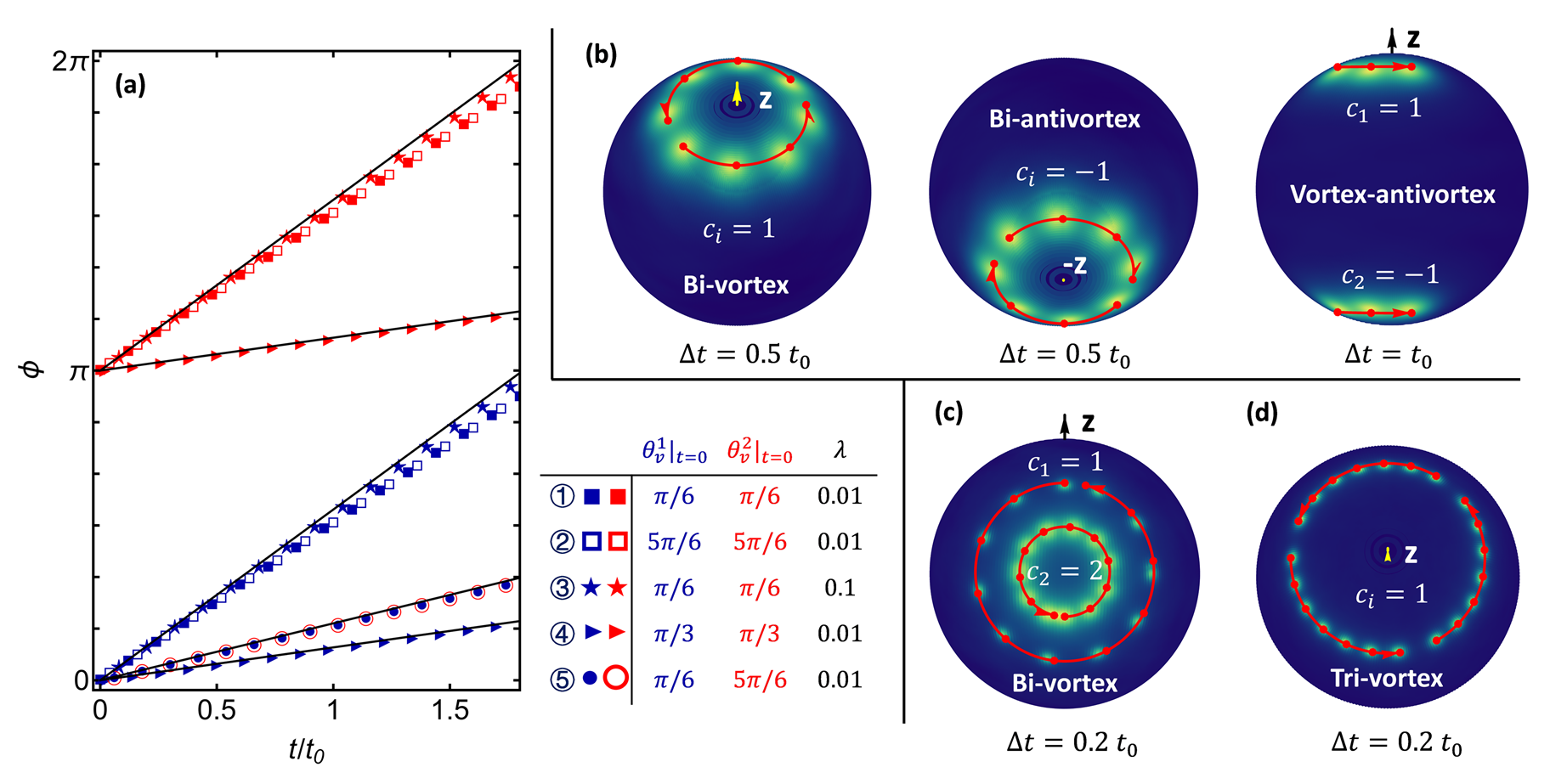}
\caption{The trajectories of vortex cores. Time unit is $ t_0 =1/\varepsilon_0$, the unit for $\lambda$ is $1/m$, and $N = 10^5$.  (a) Bi-vortex systems with $|c_i| = 1$. The table shows the initial $\theta_v^i$ of the two vortices for cases \textcircled{\scriptsize 1}-\textcircled{\scriptsize 5}. Blue (red) refers to the first (second) vortex. The filled (empty) markers refer to vortices (antivortices). All $\theta_v(t)$ almost keep constant during the evolution (not presented) and $\phi_v(t)$ increase linearly. The markers are results from GPE simulations. The black solid lines are theoretical predictions from Eq.\,\eqref{preci}. When two vortices are close (case \textcircled{\scriptsize 1}), a larger interaction (case \textcircled{\scriptsize 3}) validates the point-vortex approximation, and thus the numerical results become more consistent with the theoretical predictions. Cases \textcircled{\scriptsize 1}, \textcircled{\scriptsize 2}, and \textcircled{\scriptsize 5} have the same initial angular momenta $\bm{l}_1$ and $\bm{l}_2$, corresponding to vortex-vortex, antivortex-antivortex, and vortex-antivortex cases. (b) shows the vortex trajectories on sphere for these three cases. The red lines with arrows are theoretical predictions, where points refers to vortex positions at different time and $\Delta t$ is the time separation for neighbors. The background colors are layer superpositions of atom density from GPE simulation at different times.  (c) Bi-vortex system with high winding numbers. (d) Tri-vortex system.
}
\label{fig3t}
\end{figure*}


To examine the above theoretical prediction, we utilize GPE \eqref{gpe} to simulate the BEC dynamics. The initial state is prepared by imaginary time evolution of the GPE with phase imprinting, where the phase profile is given by Eq.\,\eqref{twovort}. The method allows for numerical creation of quantized vortices with well-defined winding numbers at specified locations on the sphere.

For two-vortex systems, from the precession equation, we find
$\dot{\bm l}_1 = \beta {\bm l}_2 \times {\bm l_1}$ and $\dot{\bm l}_2 = \beta {\bm l}_1 \times {\bm l_2}$, where $\beta = \varepsilon_0 /(1- 4 {\bm l}_1 \cdot {\bm l}_2/ c_1 c_2)$. The two spins precess with each other. Moreover, the total spin ($\bm{l}_T = \bm{l}_1 + \bm{l}_2$) is conserved during the evolution, and hence
\begin{equation} \label{preci}
\dot{\bm{l}}_i(t) = \beta \bm{l}_T \times \bm{l}_i.
\end{equation}
The initial $\{\bm{l}_1, \bm{l}_2\}$ fully determines the trajectory of $\bm{l}_i(t)$. Since $\bm{l}_i = {c_i} {\textbf n}^i_v /2 $, it means that a vortex of $c_i$ at ${\textbf n}^i_v$ and an antivortex of $-c_i$ at $-{\textbf n}^i_v$ have an exactly same angular momentum. This hints a duality as shown below.

Figure\,\ref{fig3t}\,(a) presents the two-vortex dynamics for $|c_i| = 1$. In \textcircled{\footnotesize  1}, two vortices are initially positioned at $(\theta_v^1, \phi_v^1) = (\pi/6, 0)$ and $(\theta_v^2, \phi_v^2) = (\pi/6, \pi)$, and they move around each other: $\theta_v^i(t)$ remains constant (not shown) and $\phi_v^i(t)$ increases linearly as predicted. The same results are also observed for two antivortices initially placed at positions opposite to these two vortices (\textcircled{\footnotesize 2} with $\theta_v^i=5\pi/6$). The prediction becomes more accurate when increasing initial vortex separation (\textcircled{\footnotesize 4} with $\theta_v^i = \pi/3$) or the interatomic interaction (\textcircled{\footnotesize 3}), where the point-vortex approximation is more applicable. When one vortex in \textcircled{\footnotesize 1} is replaced by an antivortex at the corresponding opposite position (\textcircled{\footnotesize 5}), the vortex-antivortex pair move in parallel \cite{Sun2020,Bereta2021}. Figure\,\ref{fig3t}\,(b) plots the vortex trajectories on the sphere for \textcircled{\footnotesize 1}, \textcircled{\footnotesize 2}, and \textcircled{\footnotesize 5}, all with the same initial  $\{\bm{l}_1, \bm{l}_2\}$. These three systems share the same group of precession equations \eqref{preci}, where $\beta$ depends on the sign of $c_1 c_2$. This illustrates the duality among vortex-vortex, antivortex-antivortex, and vortex-antivortex systems with different monopoles, regardless of the magnitude of the precession frequency. This duality also works for cases with high winding numbers.

Figure\,\ref{fig3t}\,(c) presents two-vortex dynamics with $c_1 = 1$ and $c_2 = 2$, where the vortex trajectories closely follow analytic prediction. The vortex with $c_2 = 2$ maintains remarkable stability during the first orbital period. However, since the vortex core size increases with $c_i$ \cite{Pethick2008}, it requires stronger interatomic interaction for accurate predictions within the point-vortex approximation.

For multi-vortices, an impressive result is the conservation of the total angular momentum ($\bm{l}_T = \sum_i \bm{l}_i$) during the evolution. In Fig.\,\ref{fig3t}\,(d), we consider three vortices of $c_i = 1$ placed equidistantly. The vortices rotate along the axis of $\bm{l}_T$ periodically. For general multi-vortices, the system enter a regime of dynamical chaos --- the trajectory evolution lacks periodicity and becomes sensitive to initial positions. However, the early-time dynamics are consistent with our theoretical prediction, demonstrating the validity of composite particle picture.


{\it Geometry-Gauge Interplay Mechanism.} So far, we have developed the kinetic theory of vortices in BEC on a sphere. The vortex-monopole index theorem guarantees the construction of composite topological particles. The vortex motion is exactly described by the precession of the composite's spin, which is a reflection of the spherical geometry. The effect of the magnetic monopole is to alter the vortex distribution through the index theorem, which is the topological characteristics of a closed manifold. That means, given a vortex distribution $\{c_1, \cdots, c_n\}$, increasing the monopole charge by $\Delta g$ is equivalent to add new vortices $\{c_{n+1}, \cdots\}$ with $\sum_{i>n} c_i = 2\Delta g$ to the sphere system. These newly added vortices dynamically evolve with the existing vortices, instead of  providing a static background field without feedback. This first reveals the dynamical interplay between geometry and magnetic fields in the BEC system.

Note that the effective magnetic field can be simulated by the Berry curvature in cold atom systems \cite{Zheng2025,Dalibard2011,Niu2003}. The results indeed show the joint effects of different kinds of curvatures. Regarding the current research on BKT transitions \cite{Tononi2019,Tononi2022} and turbulence \cite{Kanai2021,Yu2025,Dritschel2015} on the sphere in the absence of a magnetic monopole, our study will also stimulate further research on the monopole effects on such topological systems with net vorticity.

\begin{acknowledgments}
We acknowledge the discussion from Congjun Wu, Lichen Zhao, Shanchao Zhang, Biao Wu, and Wu-Ming Liu. This work is supported by the National Natural Science Foundation of China under grants No.\,12105223, No.\,12175180, No.\,11934015, and No.\,12247103, the Major Basic Research Program of Natural Science of Shaanxi Province under grants No.\,2017KCT-12 and No.\,2017ZDJC-32,  Shaanxi Fundamental Science Research Project for Mathematics and Physics under grant No.\,22JSZ005 and No.\,22JSQ041, and Natural Science Basic Research Program of Shaanxi under Grant No.\,2024JC-YBMS-022. This research is also supported by the Youth Innovation Team of Shannxi Universities and The Double First-class University Construction Project of Northwest University.
\end{acknowledgments}

\appendix
\section{End Matter}

\begin{figure}[h]
\includegraphics[width=\columnwidth]{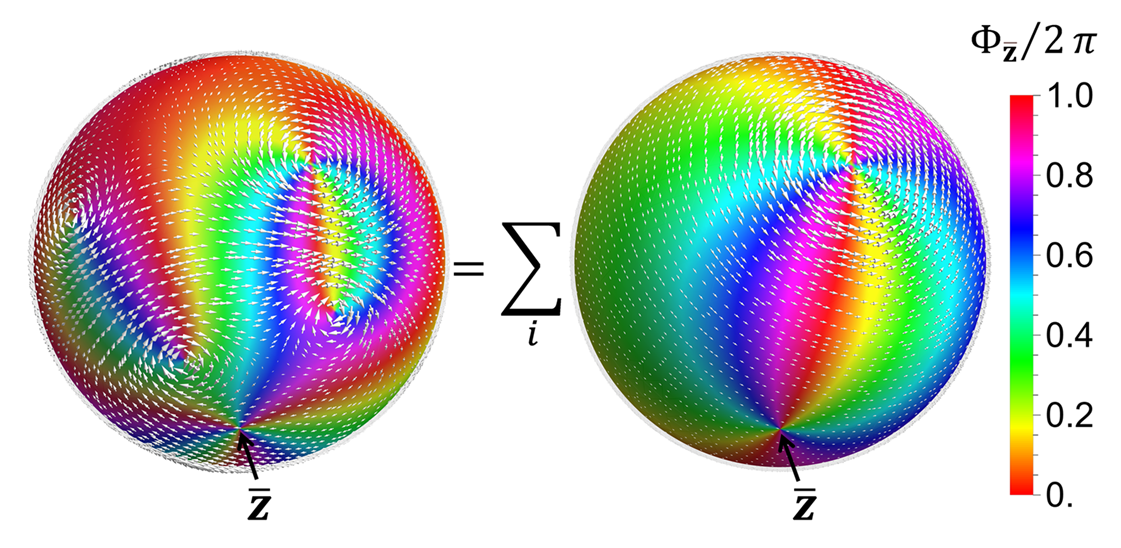}
\caption{A BEC vortex system with winding numbers $\{c_1, c_2, \cdots, c_n \}$ on a sphere is equivalent to an interacting system of composite topological particles, each of which consists of a vortex and a matching monopole. This equivalence also holds for the special case without a monopole, where $g=0$. Here, the monopole magnetic charge is $g=3$. The winding number is zero at the south pole, even though the phase
accumulation surrounding the south pole is $-6\pi$ (left one) in the $\bar{\textbf{z}}$-gauge.}
\label{fig0}
\end{figure}

In Fig.\,\ref{fig0}, we plot the velocity distribution for typical vortex systems on the Haldane sphere.

\begin{figure}[h]
\includegraphics[angle=0,width=\columnwidth]{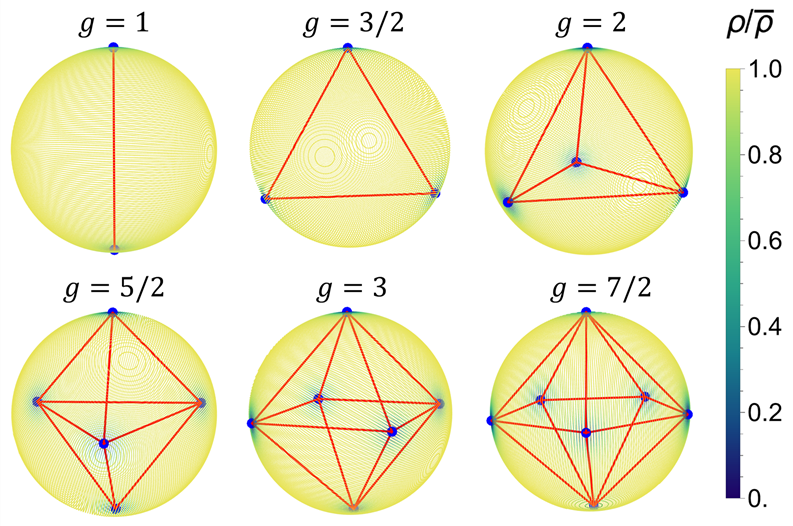}
\caption{Particle density distribution for the ground state obtained by imaginary time evolution of the GPE. Blue points (ends of red lines) are obtained by minimizing $E_{\text{int}}$ for different $g$. Here $\lambda = 0.01/m$ and $N = 10^{5}$.}
\label{fig4t}
\end{figure}

In Fig.\,\ref{fig4t}, we determine the vortex lattice by imaginary time evolution of the GPE, the result of which are consistent with that obtained by  minimizing the interaction energy \eqref{eq:interaction}. For integer $g$, the vortex lattice patterns with weak interatomic interaction have been obtained in Ref.\,\cite{Zhou2018}. For a strong interatomic interaction in our case, the lattice patterns are similar. In half-integer cases, for $g = 3/2$, there are three vortices forming an equilateral triangle on the equatorial plane, while for $g = 5/2$ and $7/2$, besides the two vortices located at the north and south poles, the others form an equilateral triangle and a regular pentagon on the equator, respectively.  We only present cases with $g \leq 7/2$, since these monopoles of small magnetic charge should be achievable in current experiments \cite{Zheng2025}.

\newpage
\onecolumngrid

\appendix

~\\

{\centering \Large \textbf{Supplementary Materials}}

~\\
In this supplementary, we present the details of the derivation for the gauge connection $\xi_{{\textbf n} \bar{\textbf z}} $, the velocity field in a single vortex-monopole composite, the interaction between composites,  the equation of motion for vortices, and the numerical method.

\section{The explicit form of $\xi_{\textbf n\bar{\textbf z}} $}
In this section, we determine the gauge connection $\xi_{\textbf n\bar{\textbf z}} $ from the relation $\nabla_s \xi_{\textbf n\bar{\textbf z}} = \textbf{A}_{\textbf n} - \textbf{A}_{\bar{\textbf z}}$ through integration along a special path on the sphere.

First, we consider the path integral along fixed $\phi = \phi_{\textbf n}$:
\begin{equation}
\int_{\phi=\phi_{\textbf n}} (\textbf{A}_{\textbf n} - \textbf{A}_{\bar{\textbf z}})\cdot d\bm \ell = R \int_{\phi=\phi_{\textbf n}} (\textbf{A}_{\textbf n} - \textbf{A}_{\bar{\textbf z}})\cdot \hat{\textbf e}_\theta d\theta,
\end{equation}
Note that
\begin{equation}
\textbf{A}_{\textbf m} \cdot \hat{\textbf e}_\theta \propto {\textbf m \times \textbf r \cdot \hat{\textbf e}_\theta} = {\textbf m \cdot \hat{\textbf e}_\phi} =  -\sin\theta_{\textbf m} \sin(\phi - \phi_{\textbf m}).
\end{equation}
Since $\phi =\phi_{\textbf n}$, we have
\begin{equation}
\textbf{A}_{\textbf n}(\textbf r) \cdot \hat{\textbf e}_\theta {\big |}_{\phi =\phi_{\textbf n}} =0.
\end{equation}
For $\textbf{A}_{\bar{\textbf z}}$, since $\sin\theta_{\bar{\textbf z}} =0$, we have
\begin{equation}
\textbf{A}_{\bar{\textbf z}}(\textbf r) \cdot \hat{\textbf e}_\theta {\big |}_{\phi =\phi_{\textbf n}} =0.
\end{equation}
These results lead to the integral along fixed $\phi = \phi_{\textbf n}$ vanishing.

Next, we consider integrating along a latitude circle with fixed $\theta$. The path integral becomes:
\begin{equation}
\xi_{\textbf n\bar{ \textbf z}} \sim \int_{\phi_{\textbf n}}^\phi (\textbf{A}_{\textbf n} - \textbf{A}_{\bar{\textbf z}})\cdot \hat{\textbf{e}}_\phi ~ R\sin\theta d \phi,
\end{equation}
where
\begin{equation}
(\textbf{A}_{\textbf n} - \textbf{A}_{\bar{\textbf z}})\cdot \hat{\textbf{e}}_\phi = \frac{g\left[\sin\theta_{\textbf n}\cos\theta\cos(\phi - \phi_{\textbf n}) - \cos\theta_{\textbf n}\sin\theta\right]}{R\left[1 - \sin\theta_{\textbf n}\sin\theta\cos(\phi - \phi_{\textbf n})- \cos\theta_{\textbf n}\cos\theta\right]} - \frac{g \sin\theta}{R (1 + \cos\theta)}.
\end{equation}
After careful evaluation, we obtain:
\begin{equation}
\xi_{\textbf n\bar{ \textbf z}} \sim 2g\left[ -\frac{\phi - \phi_{\textbf n}}{2} - \text{Arctan}\left( \frac{\sin\left(\frac{\theta+\theta_{\textbf n}}{2}\right)\sin\left(\frac{\phi-\phi_{\textbf n}}{2}\right)}{\sin\left(\frac{\theta-\theta_{\textbf n}}{2}\right)\cos\left(\frac{\phi-\phi_{\textbf n}}{2}\right)} \right) \right] + \text{const.}
\end{equation}
The explicit transformation to the $\bar{ \textbf z}$-gauge can be written as
\begin{align}\label{connection}
  \xi_{ \textbf n  \bar{ \textbf z}}(\textbf r) &= 2 g \, \text{Arg} \Bigg\{ \exp \left(-i \frac{\phi - \phi_{\textbf n}}{2} \right) \bigg[-\sin \left(\frac{\theta-{\theta_{\textbf n}}}{2}\right) \notag\\
  &\cos \left(\frac{\phi -{\phi_{\textbf n}}}{2}\right) + i \sin \left(\frac{\theta +{\theta_{ \textbf n}}}{2}\right) \sin \left(\frac{\phi -{\phi_{\textbf n}}}{2}\right)\bigg]\Bigg\}.
\end{align}
In our convention, for the north and south pole points, we set $\phi_{\textbf n} = 0$. Then, we have $\xi_{ \bar{\textbf z} \bar{\textbf z}}(\textbf r) =0$ and $\xi_{\textbf z\bar{\textbf z}}(\textbf r) = -2g(\phi+\pi)$, after discarding a multiple of $2\pi$.

\section{The velocity field in a single vortex-monopole composite}

For a general vortex position $R \textbf{n}_v$ on the sphere, the phase in the gauge $\bar{\textbf{n}}_v \equiv -{\textbf{n}}_v$ is given by $\Phi^c_{\bar{\textbf n}_v}(\textbf{r}, \textbf{n}_v) = c \phi'$, where $\phi'$ is the azimuthal angle in a coordinate system with ${\textbf{n}}_v$ as the $z'$-axis. Explicitly,
\begin{eqnarray}
 \phi' &=&  \text{Arg} \left\{ \textbf{r} \cdot [(\textbf{z} \times \textbf{n}_v) \times \textbf{n}_v] + i \textbf{r} \cdot (\textbf{z} \times \textbf{n}_v) \right\} \notag \\
 &=& \text{Arg} \big\{\sin(\theta_v ) \big[\sin (\theta ) \cos (\theta_v) \cos (\phi -\phi_v)-\cos (\theta ) \sin (\theta_v)
 +i \sin (\theta ) \sin (\phi -\phi_v)\big]\big\} \notag \\
&=&
\text{Arg} \big[\sin (\theta ) \cos (\theta_v) \cos (\phi -\phi_v)-\cos (\theta ) \sin (\theta_v)  +i \sin (\theta ) \sin (\phi -\phi_v)\big].
\end{eqnarray}

For the vortex at the north pole, the velocity distribution is \cite{Zheng2025}
\begin{equation}
  \textbf v_c (\textbf{r}; \textbf{z}) = \frac{c}{2 m R}\frac{(1+\cos{\theta})}{ \sin{\theta}}\textbf {e}_\phi,
\end{equation}
To generalize the form to arbitrary vortex position  $R \textbf{n}_v$, we use replacements $\cos\theta = \textbf{n}_v \cdot \hat{\textbf{e}}_r$, $\textbf {e}_\phi = \textbf{n}_v \times \hat{\textbf{e}}_r /\sin \theta$ and $\sin^2\theta = 1- \cos^2\theta = 1-(\textbf{n}_v \cdot \hat{\textbf{e}}_r)^2$. Finally, we have
\begin{equation}
\textbf{v}_c(\textbf{r}; \textbf{n}_v) \equiv \frac{\nabla \Phi^c_{\bar{\textbf n}_v}- \textbf{A}_{\bar{\textbf n}_v}}{m} = \frac{c}{2 m R} \frac{\textbf{n}_v \times \textbf{r}}{R - \textbf{n}_v \cdot \textbf{r}},
\end{equation}

\section{The interaction between composites}

For two composites with velocity fields $\textbf v_{c_1}$ and $\textbf v_{c_2}$, due to the rotational symmetry, we can assume that one vortex is at $(\theta_v^1, \phi_v^1) = (0, 0)$ and the other one at $(\theta_v^2, \phi_v^2) = (\theta_v, 0)$, without loss of generality. From the formula of the velocity, we have the cross term,
\begin{equation}
E_{2v} = {\rho} m R^{2} \int d\Omega  \textbf v_{c_1} \cdot \textbf v_{c_2} = \frac{\rho c_1 c_2}{4m} \int d\Omega  \frac{1+\cos{\theta}}{\sin{\theta}} \frac{\textbf {e}_\phi \cdot (\textbf{n}_v \times \hat{\textbf{e}}_r)}{ 1 -\textbf{n}_v \cdot \hat{\textbf{e}}_r}.
\end{equation}
Using the fact that $\textbf {e}_\phi \cdot (\textbf{n}_v \times \hat{\textbf{e}}_r) = - \textbf {e}_\theta \cdot \textbf{n}_v$, we obtain
\begin{eqnarray}
    E_{2v} = \frac{\rho c_1 c_2}{4m} \int d\theta d\phi \Bigg\{(1+\cos{\theta}) \frac{-\cos\theta\sin\theta_v\cos\phi  + \sin\theta\cos\theta_v}{ 1 - \sin\theta \sin\theta_v \cos\phi - \cos\theta \cos\theta_v } \Bigg\}.
    \end{eqnarray}
Partial integration gives
\begin{equation}\label{x}
 E_{2v} =
      \frac{\rho c_1 c_2}{4m} \Bigg\{\int d\phi \Big[({1+\cos{\theta}}) f(\theta,\phi,\theta_v) \Big]_{\theta=0}^{\theta=\pi} + I_2 \Bigg\},
\end{equation}
where
\begin{eqnarray}
f(\theta,\phi,\theta_v) &=& \ln(1 - \sin\theta \sin\theta_v \cos\phi - \cos\theta \cos\theta_v )  \
\end{eqnarray}
and
\begin{eqnarray}
I_2 &=& \int d\theta d\phi {\sin\theta} f(\theta,\phi,\theta_v)  = \int d\Omega \ln(1-\hat{\textbf{e}}_r \cdot \textbf n_v) .
\end{eqnarray}
Note that the integral $I_2$ does not depend on the orientation of $\textbf n_v$. Therefore, we can set $\textbf n_v = \textbf z$ in the integral $I_2$, and we get $I_2 = \int d\Omega \ln(1-\cos\theta) = 4 \pi (\ln 2-1)$. Directly calculating the first part of \eqref{x}, we finally obtain
\begin{equation}
E_{2v} = -\frac{\pi {\rho} c_1 c_2}{m} \ln(1 -\cos\theta_v) + \frac{ {\rho} c_1 c_2}{4 m} I_2.
\end{equation}
After discarding the constant term $I_2$, we obtain the formula in the main text.

\section{The equation of motion for vortices}
In the $\bar{\textbf z}$-gauge, we define
\begin{equation}
  \psi_{\bar{\textbf z}}(\textbf r)= \hat {\mathcal{R}}_z \hat {\mathcal{R}}_y \exp[i c \phi(\textbf r)] \equiv \exp[i \Phi(\textbf r)] ,
\end{equation}
and
\begin{equation}
   \hat {\mathcal{R}}_z = \exp[-i\hat{L}_{z} (g) \phi_v],~~~~  \hat {\mathcal{R}}_y = \exp[-i\hat{L}_{y} (g) \theta_v],~~~  \hat {\mathcal{R}} = \hat {\mathcal{R}}_z \hat {\mathcal{R}}_y.
\end{equation}
In Dirac notation, $|\psi_{\bar{\textbf z}} ({\textbf{n}_v}) \rangle = \hat {\mathcal{R}} |\psi_{\bar{\textbf z}}(\textbf{z})\rangle  $, where $|\psi_{\bar{\textbf z}}(\textbf{z})\rangle $ is the state with a vortex of winding number $c$ at the north pole. The angular momentum operators satisfy the commutation relations $[\hat L_i,\hat L_j] = i\hbar \epsilon_{ijk} \hat L_k$. Thus, we have
\begin{equation} \label{cccc}
\hat{\mathcal{R}}^\dagger \hat{L}_z \hat{\mathcal{R}} = -\sin\theta_v \hat{L}_x + \cos\theta_v \hat{L}_z, ~~~~ \hat{\mathcal{R}}_y^\dagger \hat{L}_y \hat{\mathcal{R}}_y = \hat{L}_y.
\end{equation}

We consider the wavefunction
\begin{equation}
  \Psi_{\bar{\textbf {z}}}(\textbf{r},t) =  \exp(-i \mathcal{E} t) \sqrt{\rho}  \prod_{j} \big[ \hat{\mathcal{R}}(g_j;\phi_v^j,\theta_v^j) \exp(i c_j \phi)\big],
\end{equation}
where $\hat{\mathcal{R}}(g_j;\theta_v^j,\phi_v^j) = \exp[-i\hat{L}_{z}(g_j) \phi_v^j] \exp[-i\hat{L}_{y}(g_j) \theta_v^j]$ rotates a vortex from the north pole to the real-time position $(\theta_v^j(t),\phi_v^j(t))$.
Note that
\begin{eqnarray}
   R^2 \int d\Omega [i\Psi^*_{\bar{\textbf {z}}} \partial_t \Psi_{\bar{\textbf {z}}}] &=&  \mathcal{E} N + \sum_j R^2 \rho \int d\Omega i \big[\hat{\mathcal{R}}(g_j) \exp(i c_j \phi)\big]^* \partial_t \big[ \hat{\mathcal{R}}(g_j) \exp(i c_j \phi )\big] \notag \\
   &=&  \mathcal{E} N + i N \sum_j  \langle \hat{\mathcal{R}}(g_j)\psi_{\bar{\textbf z}}(\textbf{z}) | \partial_t |\hat{\mathcal{R}}(g_j)\psi_{\bar{\textbf z}}(\textbf{z})\rangle \notag \\
   &=&  \mathcal{E} N + i N \sum_j  \langle \psi_{\bar{\textbf z}}(\textbf{z}) | [ \hat{\mathcal{R}}^\dagger(g_j) \partial_t \hat{\mathcal{R}}(g_j)] |\psi_{\bar{\textbf z}}(\textbf{z})\rangle
\end{eqnarray}
and
\begin{eqnarray}
 i \hat{\mathcal{R}}^\dagger(g_j) \partial_t \hat{\mathcal{R}}(g_j)
    &=& \dot{\phi}_v^j \hat{\mathcal{R}}^\dagger(g_j) \hat{L}_{z} \hat{\mathcal{R}}(g_j)  +  \dot{\theta}_v^j \hat{\mathcal{R}}_y^\dagger(g_j) \hat{L}_{y} \hat{\mathcal{R}}_y(g_j).
\end{eqnarray}
Using Eq.\,\eqref{cccc} and $\langle \psi_{\bar{\textbf z}}(\textbf{z}) | \hat{\textbf {L}}(g_j)|\psi_{\bar{\textbf z}}(\textbf{z})\rangle = \hat{\textbf e}_z c_j/2  $, we have
\begin{eqnarray}
   R^2 \int d\Omega [i\Psi^*_{\bar{\textbf {z}}} \partial_t \Psi_{\bar{\textbf {z}}}] =  \mathcal{E} N +  \frac{N }{2} \sum_j c_j  \dot{\phi}_v^j \cos\theta^j_v.
\end{eqnarray}
As a result,
 \begin{equation}
\mathcal{L} (\theta_v^1,\phi_v^1, \cdots, \dot{\theta}_v^1,\dot{\phi}_v^1, \cdots) = N \sum_i \frac{c_i}{2} \dot{\phi}_v^i\cos\theta_v^i - E_{\text{int}},
\end{equation}
after getting rid of a constant.

Minimizing the action $S = \int \mathcal{L} dt$ leads to
\begin{eqnarray}
- N \frac{c_i}{2} \dot{\phi}_v^i \sin \theta_v^i &=& \frac{\partial E_{\text{int}}}{\partial \theta_v^i} ~=~  \frac{\partial \bm l_i}{ \partial \theta_v^i} \cdot \nabla_{ \bm l_i} E_{\text{int}},  \\
N \frac{c_i}{2}
\dot{\theta}_v^i \sin \theta_v^i &=& \frac{\partial E_{\text{int}}}{\partial \phi_v^i} ~=~ \frac{\partial \bm l_i}{ \partial \phi_v^i}  \cdot \nabla_{ \bm l_i} E_{\text{int}}.
\end{eqnarray}
where $\bm l_ i \equiv c_i \textbf{n}^i_v /2$ and $\textbf{n}^i_v  = (\sin \theta_v^i \cos \phi_v^i, \sin \theta_v^i \sin \phi_v^i, \cos \theta_v^i)$.
Using the differential relation $\dot{\bm l_i} = \dot{\theta}_v^i {\partial \bm l_i}/{\partial \theta_v^i} +\dot{\phi}_v^i {\partial\bm l_i}/{\partial \phi_v^i} $, we finally get the evolution of the `spin' of composite,
\begin{equation}\label{eomx}
\dot{\bm l_i} = \frac{2}{N c_i} \nabla_{ \bm l_i} E_{\text{int}}\times \left[ \frac{\partial\bm l_i}{\partial \theta_v^i} \times \frac{\partial\bm l_i}{\sin\theta_v^i \partial \phi_v^i} \right] =
\bm \omega_i \times \bm l_i,
\end{equation}
where $\bm{\omega}_i \equiv \nabla_{\bm l_i} (E_{\text{int}}/N)$. In the above, we have used the fact that
\begin{equation}
   \frac{\partial \textbf{n}^i_v}{\partial \theta_v^i} \times \frac{\partial \textbf{n}^i_v}{\sin\theta_v^i \partial \phi_v^i} = \textbf{n}^i_v.
\end{equation}

\section{The numerical method}
The covariant Laplacian on a sphere with radius $ R $ in spherical coordinates $(\theta, \phi)$, in the presence of a gauge field $\mathbf{A} = (A_\theta, A_\phi)$, is given by
\begin{equation}
\nabla_A^2 \psi = \frac{1}{R^2 \sin \theta} D_\theta \left( \sin \theta \, D_\theta \psi \right) + \frac{1}{R^2 \sin^2 \theta} D_\phi^2 \psi,
\end{equation}
where $ D_\theta = \partial_\theta - i R A_\theta $ is the covariant derivative in $\theta$ and
 $ D_\phi = \partial_\phi - i R \sin \theta \, A_\phi $ is the covariant derivative in $\phi$. For the $\bar{\textbf z}$-gauge,
\begin{equation}
{A}_\theta (\textbf r)=0,~~~~\text{and}~~~~{A}_\phi (\textbf r)=\frac{g( 1-\cos{\theta})}{R \sin{\theta}}.
\end{equation}
Rewrite the covariant Laplacian using the variable $z =\cos\theta $,
\begin{equation}
\nabla_A^2 \psi = \frac{1}{R^2}\left\{\partial_z\left[(1-z^2)\partial_z\psi\right] + \frac{1}{1-z^2}\left[\partial_\phi - ig(1-z)\right]^2\psi\right\}.
\end{equation}\
The Gross-Pitaevskii equation (GPE) reads
\begin{equation}\label{xxx}
  i \partial_t \psi  = - \frac{1}{2m}\nabla_A^2 \psi + \lambda |\psi|^2\psi.
\end{equation}
Using the split-step method, we develop the Python code to time-evolve the GPE for $z \in [-1, 1]$ and $\phi \in [0, 2 \pi)$. In the numerical simulation, we discretize $z$  in to $N$ points, with  $z_i  =  -1 + (i+1/2) \Delta z$ and $\Delta z = 2/N$, where $i=0,\cdots, N-1$. The endpoints $z =-1$ and $z=1$ are not involved. We use a $256\times256$ grids for $(z,\phi)$ in our simulations.

%
\end{document}